\documentclass{aa}
\usepackage{graphics}

\usepackage{natbib}
\usepackage{graphicx}

\input{psfig}

\begin{document}

\title{X-ray emission from Saturn}

\author{J.-U. Ness\inst{1}, J.H.M.M. Schmitt\inst{1}, S.J. Wolk\inst{2}, K. Dennerl\inst{3}, V. Burwitz\inst{3}}
\institute{
Hamburger Sternwarte, Universit\" at Hamburg, Gojenbergsweg 112,
D-21029 Hamburg, Germany
\and
Chandra X-ray Center, Harvard-Smithsonian Center for Astrophysics, 60 Garden Street, Cambridge, MA 02138, USA
\and
  Max-Planck-Institut f\"ur extraterrestrische Physik (MPE), Postfach 1312, 85741 D-Garching, Germany}

\authorrunning{Ness et al.}
\titlerunning{Chandra observation of Saturn}
\offprints{J.-U. Ness}
\mail{jness@hs.uni-hamburg.de}
\date{Received \today;accepted ...}

\abstract{
We report the first unambiguous detection of X-ray emission originating from
Saturn with a Chandra observation, duration 65.5\,ksec with ACIS-S3. Beyond the
pure detection we analyze the spatial distribution of X-rays on the planetary
surface, the light curve, and some spectral properties. The detection is based
on 162\,cts extracted from the ACIS-S3 chip within the optical disk of
Saturn. We found no evidence for smaller or larger angular extent. The
expected background level is 56\,cts, i.e., the count rate is
$(1.6\,\pm\,0.2)\cdot10^{-3}$\,cts/s. The extracted photons are rather
concentrated towards the equator of the apparent disk, while both polar caps
have a relative photon deficit. The inclination angle of Saturn during the
observation was $\sim -27^\circ$, so that the northern hemisphere was not
visible during the complete observation. In addition, it was occulted
by the ring system. We found a small but significant photon excess
at one edge of the ring system. The light curve shows a small dip twice at
identical phases, but rotational modulation cannot be claimed at a significant
level. Spectral modeling results in a number of statistically, but not
necessarily physically, acceptable models. The X-ray flux level we calculate
from the best-fit spectral models is
$\sim 6.8\cdot10^{-15}\mbox{\,erg\,cm}^{-2}\mbox{\,s}^{-1}$ (in the energy
interval 0.1--2\,keV), which corresponds to an X-ray luminosity of
$\sim 8.7\cdot10^{14}\mbox{\,erg}\mbox{\,s}^{-1}$. A combination of scatter
processes of solar X-rays require a relatively high albedo favoring internal
processes, but a definitive explanation remains an open issue.
\keywords{planets and satellites: general - planets and satellites individual: Saturn - X-rays: general}
}
\maketitle

\section{Introduction}
\label{intro}

X-ray emission from solar system objects has so far been detected from the
Earth \citep{Earth,rug79,fink88}, from the Moon \citep{moon,schmitt91}, from a
number of comets \citep[e.g.,][]{lisse96,dennerl97,mumma97}, from Jupiter
\citep{metzger83}, from the Galilean satellites Io and Europa
\citep{elsner02}, from Venus \citep{venus}, Mars \citep{mars}, and marginally
from Saturn \citep{ness00}. The observed X-ray emission appears to have
different physical origins in the different objects. The principal X-ray
production mechanism for Moon, Earth, Venus, and Mars is scattering of solar
X-rays. Auroral X-ray emission has been found from the Earth and from
Jupiter, and similar emission from the outer planets is anticipated.

Aurorae on Earth and Jupiter are generated by charged particles precipitating
into the planetary atmosphere along the magnetic field lines. While at Earth the
precipitating flux consists of solar wind electrons, {\it Einstein}
observations were interpreted in a way that the Jovian X-rays are caused by
heavy ion precipitation with oxygen and sulfur ions originating from the
volcanically active moon Io \citep[e.g.,][]{metzger83}. Further support for this
scenario came
from a direct observation of heavy ions in Jupiter's magnetosphere with the
{\it Voyager} spacecraft and from a comparison of ROSAT observations in the
soft X-ray spectrum with model-generated bremsstrahlung and line emission
spectra \citep{waite94}. In an analysis of O\,{\sc i} ($\lambda$1304) and
S\,{\sc ii} ($\lambda$1256) measured with HST, \cite{trafton} found only upper
limits, but they note that these upper limits were still consistent with the
existence of sufficient heavy ions among the precipitating particles to explain
the X-ray observations. However, a recent Chandra HRC observation of Jupiter
carried out in December 2000
for an entire 10 hour rotation \citep{waite01,gladst01} has put serious doubt
on this theory. The X-ray emission in their high-resolution image is found to
be concentrated near the magnetic poles, and a peculiar 45 minute pulsation
similar to high-latitude radio pulsations previously detected by the
Galileo and Cassini spacecraft was found. The production of X-ray emission
concentrated so close to the poles cannot be explained by ions coming from near
Io's orbit. The polar emission was identified to be stronger at the north pole
than in the south polar region \citep{waite96}. Also, equatorial emission was
identified at a low level, probably originating from different scatter
mechanisms of solar X-rays. An analysis of Chandra observations with ACIS
from November 1999 also showed evidence for
soft X-ray emission from the Galilean satellites Io and Europa, and probably
Ganymede \citep{elsner02}. They interpret the emission as a result of
bombardment of their surfaces by energetic ($>10$\,keV) H, O, and S ions
originating from the region of the Io Plasma Torus (IPT). The IPT itself
was found to emit soft X-rays, which appears at the low end of the ACIS-S3
energy band, but \cite{elsner02} found an unresolved line or line complex
indicative of oxygen.

\subsection{Previous X-ray observations of Saturn}

Saturn was observed with the {\it Einstein} Observatory IPC for about
10~ksec, but no X-ray emission was detected, leading \cite{gilman86} to the
conclusion that instead of heavy ion precipitation, electron bremsstrahlung was
the more likely X-ray production mechanism for Saturn. With this assumption they
calculated a $3\sigma$ upper limit for the Saturnian
X-ray flux at Earth of $1.7\cdot10^{-13}\mbox{\,erg\,cm}^{-2}\mbox{\,s}^{-1}$, a
value consistent with an expected energy flux at Earth of
$8\cdot10^{-16}\mbox{\,erg\,cm}^{-2}\mbox{\,s}^{-1}$, obtained from a model
calculation by \cite{gilman86} based on UV observations \citep{sandel82} and
the assumption of thick-target bremsstrahlung at high latitudes.\\
A marginal X-ray detection of Saturn was obtained in a systematic analysis of
ROSAT PSPC data on trans-Jovian planets by \cite{ness00}, although no detection
was expected from auroral thick target bremsstrahlung models. In a 5349 seconds
PSPC observation 22 counts were recorded in a box centered at the position of
Saturn while only 7.6 counts were expected from background. The probability of
measuring 22 counts or more with only 7.6 counts being expected is
$1.7\cdot10^{-5}$, assuming Poisson statistics, hence the formal significance
of the detection is quite high. The 14.4 counts, formally attributed to Saturn
correspond to an energy flux of
$1.9\cdot10^{-14}\mbox{\,erg\,cm}^{-2}\mbox{\,s}^{-1}$ using a conversion
factor of $6\cdot10^{-12}\mbox{\,erg\,cm}^{-2}\mbox{\,s}^{-1}$.
This value is significantly higher than the model calculation by \cite{gilman86}
but it is not in contradiction to the upper limit estimated from the
{\it Einstein} observation. In any case, Saturn is not as X-ray bright as
Jupiter, and little could be inferred about the X-ray spectrum of Saturn, except
that it should be very soft since no detection in the hard ROSAT band was
obtained. Also, nothing could be derived about the spatial location of the
Saturnian X-ray source because of the low counting statistics and the low
angular resolution of the ROSAT PSPC.
A recent observation of Saturn with XMM-Newton is presented by \cite{xmmsat}
and their results compare very well with our Chandra observations.\\
We present a new observation of Saturn carried out with Chandra. With the high
spatial resolution of the ACIS detectors, we choose an exposure time of
$\sim 70$\,ksec, sufficiently high to detect X-ray emission even if, in a
worst case, the emission was randomly distributed over the planetary surface.
The observation setup and data analysis is described in Sect.~\ref{anal},
the results are described in Sect.~\ref{results} and discussed in
Sect.~\ref{disc}, and our conclusions are presented in Sect.~\ref{concl}.

\section{Observations and Data Analysis}
\label{anal}

\subsection{Observations}

In April 2003, we obtained an X-ray observation of Saturn with ACIS-S3 aboard
Chandra for a total of 71.5\,ksec, but Saturn was actually only 65.5\,ksec
in the field of view (see Table~\ref{tab1}). We used the
back-illuminated S3 chip in order to take
advantage of this CCD's sensitivity to low energy X-rays. The primary
concern of the observational setup was loading the CCD due to optical
light from Saturn. Scaling from earlier observations of Jupiter, 
we expected an optical load of about 8 ADU/pixel per 3.2 second ACIS frame.
While this is below the event split event definition and would not by
itself create false events, it would strongly bias the data toward higher
energies. We used the 1/4 subarray mode to reduce the frame time and thus
lower the expected optical loading to about 2 ADU/pixel/frame. In order to
measure and correct for the residual loading of the CCD due to optical
light from Saturn we used the ``Very Faint'' (VF) telemetry format.
The telemetry saturation limit in this mode is 68.8 events per
second. We found on average 5 events per second and clearly lost no frames.
The spacecraft was re-pointed after the first 35\,ksec to follow the planet's
proper motion. Each pointing was oriented so that the planetary motion
went from one extrema to the other of the subarray traveling 
along the CCD's node one, perpendicular to the long axis of the
subarray.

\begin{table}[!ht]
 \caption{\label{tab1}Overvation details for Saturn.}
  \begin{tabular}{lr}
   ObsID & 3725 / 4433 \\
   Exp. time & 71.5\,ksec\\
   On-time   & 65.5\,ksec\\
   Start Time & 2003-04-14 07:53\\
   Stop Time & 2003-04-15 04:17\\
   RA$^a$ & 05$^h$35$^m$46$^s$\ --\ 05$^h$36$^m$06$^s$\\
   DEC$^a$ & 22$^\circ$21\arcmin 57\arcsec\ --\ 22$^\circ$22\arcmin 30\arcsec\\
   Angular diam. & 17.5\,\arcsec\\
   distance (Earth) & 9.5\,AU\\
   distance (Sun) & 9.0\,AU\\
   inclination  & -27$^\circ$\\
\hline
  \end{tabular}
\\
$^a$Coordinates as seen from Chandra (at start and stop times)
\end{table}

The observation setup was chosen to prevent optical loading. From the photon
events (Fig.~\ref{pos}) no indication for the rings can be recognized, which
would have shown up in the case of optical loading. We therefore conclude
that the setup was successful. Nevertheless we applied an energy correction
scheme ``biasevt1.pl'' developed by Peter G. Ford from MIT.
This perl script corrects the nominal 3x3 event island by using
information contained in the 5x5-pixel event island telemetered in the
Very Faint Mode. The typical region of influence of an
X-ray event is limited to a central pixel and adjacent pixels.
Members of the event island separated from the central pixel by an
intermediate pixel provide information on the local background.
For each event, biasevt1.pl subtracts the mean PHA value calculated from the
outer 16 pixels from the PHA of each of the inner nine pixels.
Filters are applied to reject background which is contaminated by
an X-ray event. This simulates what the PHA values would have been
in the absence of optical loading. The mean shift per
pixel in the 3x3 event island was 1.14 ADU, consistent with expectations.
The data are then reprocessed with the CIAO tool acis\_process\_events to
recalculate the event's true total pulse height after Charge Transfer
Inefficiency (CTI) corrections are made. Finally acis\_process\_events
applied a gain correction to each event to determine the final energy.
The pixel coordinates listed in the processed data file are converted into
sky-centered RA and DEC positions using the CIAO dmcopy command.
In order to test for the effects from the energy correction scheme,
we compared the spectra obtained from the corrected dataset and
from the non-corrected dataset. We found the two spectra to be practically
identical as expected from the small ($<$ 2 ADU) spectral shift.

\subsection{How to find Saturn's X-ray photons}
\label{extraction}

The photon positions on the original sky-centered image (in RA/DEC
coordinates) show no trace of Saturn nor can any strong background sources be
identified. We calculated an expected path using ephemeris data from the
JPL/SSD ephemeris
generator\footnote{http://ssd.jpl.nasa.gov/cgi-bin/eph} and orbit data of the
satellite provided in the orbit file (which is part of the ephemeris products
delivered with the observation). The coordinates of Saturn at start and stop
time of the observation are given in Table~\ref{tab1}.

\begin{figure}[!ht]
  \resizebox{\hsize}{!}{\includegraphics{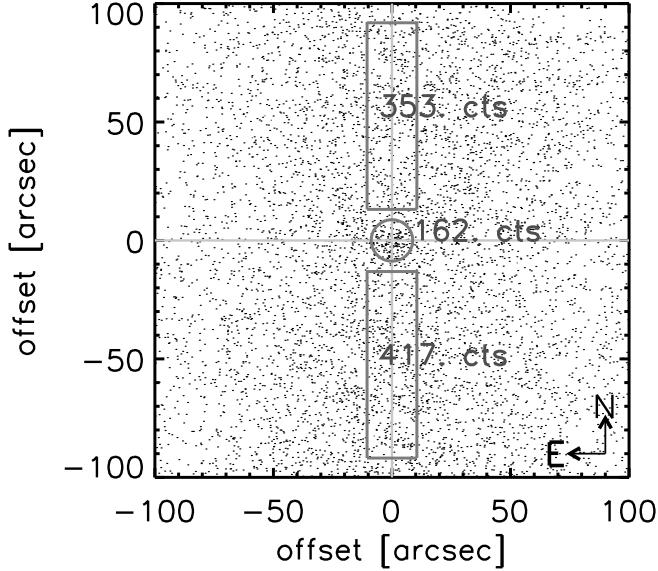}}
\caption{\label{pos}Photon events of Saturn extracted for E = 0.1--2\,keV. The
source photons are extracted from a circular region with the apparent diameter
of Saturn (17.5\arcsec, thus an extraction area of 240.528\arcsec$^2$) and the
background photons from within two 21\arcsec $\times$\,78.75\arcsec\ boxes
above and below the source (total area 3307.5\arcsec$^2$). Plotted are the
photon shifts relative to Saturn.}
\end{figure}

From the expected position of Saturn at any given time during the observation
we calculate offset coordinates with respect to the center of Saturn
from the RA and DEC coordinates for each individual photon\footnote{Since
the RA axis increases from right to left we flipped the offset positions
for a correct representation, yielding up=North and left=East}. This
transformation shifts all photons that potentially originate from Saturn to
a central ``saturnocentric'' position. We verified our extraction
procedures with the Chandra data on Jupiter, which is so strong, that
all X-ray photons originating from Jupiter can be identified without any
shifting. In Fig.~\ref{pos} the transformed photon positions are shown in
``saturnocentric'' coordinates. The extraction regions for the source (circle
with 17.5\arcsec\ diameter at origin; see Table~\ref{tab1}) and
background (large boxes above and below the source) are overlaid.
The effective exposure time is significantly smaller towards high offset
values, an effect due to the transformation. We investigated this effect by
constructing a rough exposure map consisting of nine vertical strips. For 
each strip we calculate the effective exposure time from the difference of
photon arrival times of the first and last photon. We found the effective
exposure time to vary significantly from the central strip to high offset
values. For the background extraction we therefore choose the
extraction regions near the RA position of the source and the offset mainly
in DEC direction, i.e., above and below the source. The extraction regions
are marked in Fig.~\ref{pos} and the number of photons counted in each
extraction region is given above the respective region.\\
With the chosen background method we neglect effects from background
contributions from real X-rays from behind Saturn, which will show up
in our extracted background, but are blocked by the planet in the source
extraction region. This implies an overestimation of the instrumental
background and thus an underestimation of the source flux. Background studies
at high galactic latitudes were carried out by, e.g., \cite{markev}, indicating
a low X-ray background contribution at sufficiently high galactic latitudes.
From their wide-band fluxes we estimate the source flux for Saturn to be
underestimated by at most 10\%. For conservative analysis we apply no
corrections on our instrumental background.

\subsection{Lightcurve and Spectrum}
\label{spectrum}

For the analysis of the X-ray lightcurve and the X-ray spectrum of Saturn the
photons within a circle of 10\arcsec\ radius around its nominal position were
used. This radius is somewhat larger than the 8.75\arcsec\ radius of the
Saturnian disk, to ensure that all photons from Saturn are collected, allowing
for uncertainties in the absolute attitude reconstruction and some
redistribution of the photons due to the PSF of the X-ray telescope. In order
to avoid any effect from different exposure times along the x-axis, we choose
the extraction regions for the background as vertical strips above and below
the source, just as in Fig.~\ref{pos}. With this
extraction radius we analyze 197\,counts with an expectation of 72.4\,counts
from the background. We binned the arrival times and the photon energies with
binsizes 4.6\,ksec for the light curve and 50\,eV energy bins for the spectrum.
For the spectrum we used only photons below 2\,keV (Fig.~\ref{spec}).
We counted the photons assigned to Saturn and to the background separately
(see Fig.~\ref{pos}) in each bin in order to obtain a spectrum and a lightcurve
for the source and the background. The results are shown in Figs.~\ref{lc} and
\ref{spec} and are discussed in Sect.~\ref{results}.\\
For the purpose of spectral modeling with XSPEC the raw spectrum of Saturn was
adaptively rebinned so that each bin contains at least 15 photons. The
spectral modification due to the contamination layer on ACIS was
taken into account by a multiplicative term {\em acisabs}.

\section{Results}
\label{results}

\subsection{Identification of X-rays from Saturn}

In Fig.~\ref{pos} we show the result from our transformation procedure. The
axes give the offset coordinates from the center of Saturn. From
the circular extraction region centered on Saturn's expected position we find
162 counts in the energy range 0.1--2\,keV within the known optical extent of
Saturn's disk. We estimate the number of background photons (assuming of course
a constant background level) contained in the source extraction region to be
56 counts. Obviously, this detection is highly significant. The probability of
measuring 162 counts with 56 expected from the background is zero for all
practical purposes.
Therefore the ROSAT detection of Saturn reported by \cite{ness00} is confirmed.
With these numbers and the assumption of Poissonian statistics we calculate a
net count rate of (106\,$\pm$\,12.7\,cts)/65.5\,ksec $=(1.6\,\pm\,0.2)\cdot10^{-3}$\,cts/s.

\subsection{Spatial distribution of Saturnian X-ray emission}

In order to identify Saturn as an X-ray source in our Chandra image we
assumed a spatial extent identical to the apparent optical diameter
of Saturn's disk. Having found X-ray emission from Saturn we calculated the
signal-to-noise ratio of the signal attributable to Saturn as a function of the
radius of the chosen annular extraction region. We find a relatively broad
maximum extending about 1\arcsec\ beyond Saturn's apparent optical radius,
however, given the relatively small number statistics and the broad nature of
the maximum, we conclude that there is at the moment no reason to assume an
X-ray halo extending substantially above Saturn's limb.

\begin{figure}[!ht]
  \resizebox{\hsize}{!}{\includegraphics{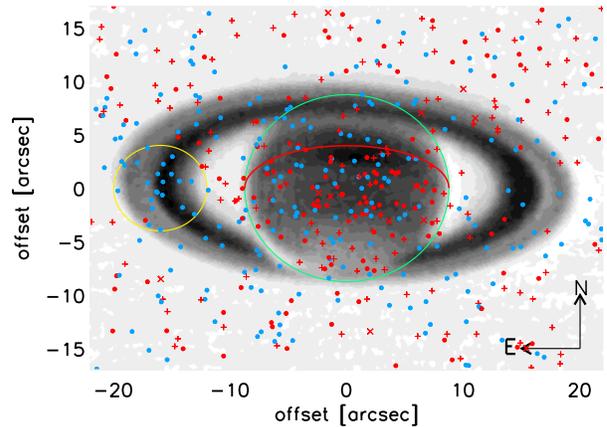}}
\caption{\label{opt}Optical image of Saturn$^2$ at the time of the Chandra
observation with the X-ray photons overlaid. The large circle marks the
extraction region with diameter 17.5\arcsec, the smaller circle indicates
and excess at the edge of the ring system. The planet was inclined by
$-27^\circ$ at the time of the observation. Symbols mark photon energies:
$<200$\,eV (x), 200--1000\,eV ($\bullet$), and $>1000$\,eV (+). Photons with
energies between 200 and 700\,eV are plotted with lighter colors.}
\end{figure}

\begin{figure}[!ht]
  \resizebox{\hsize}{!}{\includegraphics{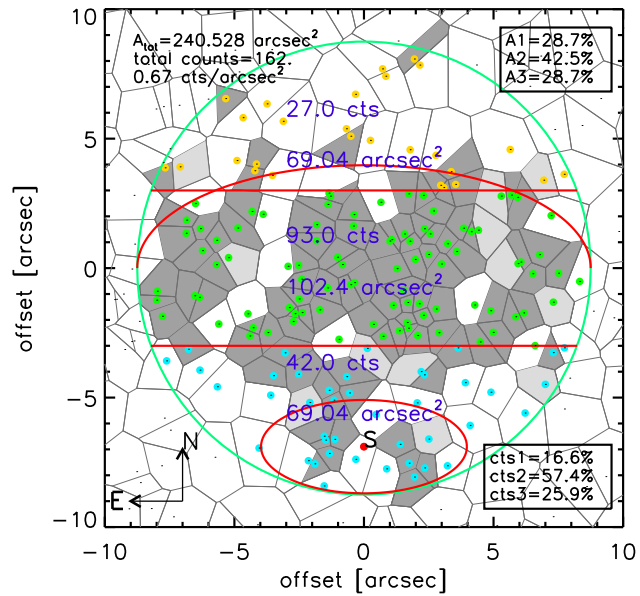}}
\caption{\label{slice}Spatial distribution of X-ray photons within
the source extraction region. The polygons are Voronoi regions comprising all
points closer to each particular photon. The 106 smallest areas are grey shaded
indicating a statistical approach to identify the net photons. Twelve more areas
are shaded with lighter grey to indicate the 1\,$\sigma$ tolerance. Three areas,
NC, EB, and SC (see text) are defined to associate polar and equatorial photons.
The geometrical equator and the visible part of the south polar region are
given by the red lines ($i=-27^\circ$).}
\end{figure}

We next investigated the spatial distribution of the recorded photons on
Saturn's apparent disk. In Fig.~\ref{opt} we plot the photons over an optical
image\footnote{Provided by {\it Bernd Flach-Wilken}, April 14, 19h\,UT, with
300mm-Schiefspiegler Feff=6m, ST237, 18\,$\times$\,0.1\,sec + Philips
ToUCam, 20\,sec} taken simultaneously with the Chandra observation. At the
time of our observation the Saturnian rings covered the northern hemisphere
potentially blocking X-ray emission originating from the northern polar
region. The planet was inclined by $-27^\circ$. The geometrical equator
is marked by a red line and the symbols representing the recorded X-ray
photons are scaled with the respective photon energies. We identify a
concentration of 500\,eV and 800\,eV photons on Saturn's disk compared to the
background. In order to investigate the homogeneity of the X-ray emission we
divided Saturn's apparent disk into three regions, a northern cap (NC) and
southern (SC) polar cap with each 28.7 \% of the total area, and an equatorial
belt (EB) containing 42.6 \% of the total area. Assuming an equal surface
brightness disk we would expect 46.5, 69.0, and 46.5 counts in NC, EB, and SC,
respectively, which has to be contrasted with 27, 93, and 42 counts actually
recorded in NC, EB, and SC, respectively. The procedure is illustrated in
Fig.~\ref{slice}.
We carried out the same analysis with the much higher SNR data available
for Jupiter and found a concentration in the polar caps compared to the
equatorial belt. Obviously, the spatial distribution of the X-ray emission in
Jupiter and Saturn is quite different. Saturn's X-ray emission is indeed
distributed inhomogeneously over its apparent disk. While for Jupiter there
is a concentration towards the poles, there is a deficit of emission from
Saturn's northern polar region, which was occulted by the ring system at
the time of our observations. However, the southern polar cap shows a deficit
as well and there is definitely a detectable concentration of the X-ray
emission towards the equator of the apparent disk.
Testing the hypothesis of a uniform distribution over the disk excluding the
northern cap we find an expected total number of uniformly distributed counts
in EB and SC of 80.6 counts in EB and 54.4 counts in SC, which has to be
contrasted with 93 and 42 counts, respectively. Under the assumption of
Poissonian statistics we calculate a total (reduced) $\chi^2_{\rm red}=2.64$.
This value allows the hypothesis of a uniform distribution to be true with a
probability of less than 10\% ($\chi^2_{\rm red}=2.99$) but more than 5\%
($\chi^2_{\rm red}=2.3$).\\
For a visual impression we constructed Voronoi areas comprising all points
closer to each particular photon. Since from the 162 photons counted within
the optical radius, 56 are statistically expected to belong to the background,
we mark the 106 smallest Voronoi polygons with grey color in Fig.~\ref{slice}
in order to identify regions of high source intensity. The next 12 smallest
areas are marked with a lighter grey to mark a 1$\sigma$ confidence level. It
can be
seen that the highest concentration is found in the center of the image. With
the inclination angle of $-27^\circ$ all emission therefore originates from
the southern hemisphere, but an enhancement of source emission towards the
south pole cannot be identified. This effect is difficult to understand in
terms of internal production mechanisms, but some scattering processes of solar
X-rays (backscattering, fluorescent scattering) could be an explanation for
the geometrical distribution of X-rays, if we assume that the rings have a
much lower X-ray albedo than the planetary disk and that they attenuate any
X-ray radiation from the planet below.\\

 From close inspection of the individual X-ray photons overlayed over the
optical image in Fig.~\ref{opt} we identify some excess in X-rays coinciding
with one edge of the ring system which cannot be seen on the other side of the
rings. We marked this area by the smaller circle at x$\sim$-16\arcsec\
and it is remarkable that the photons in this region have all roughly the same
energy. In the same fashion as for Saturn (Sect.~\ref{extraction} and
Fig.~\ref{pos}) we extract those photons within a circle of radius 4\arcsec\
(and background from boxes above and below) and
found 22\,cts with 11 expected from the background. When doing the same
extraction procedure only for photons within 200-900\,eV, the significance
is higher: still 22\,cts but with only 6\,cts expected from the background.
Interestingly an extraction in the 10\,ksec time interval 36\,ksec to 46\,ksec
(thus right after the re-pointing) returns 10\,cts with only 1.5\,cts expected
from the background. We checked the arrival times of these photons, but find
a concentration at the time right after the re-pointing of the
telescope at t$\sim 35$\,ksec not significant. We are not aware of any
consequences from the re-pointing that might lead to such a photon excess.
From the statistical point of view the detection of 22\,cts is significant,
at least within the reduced energy interval 200-900\,eV. We analyzed the
original (non-transformed) chip in this energy range and searched for
regions with 22 photons or more counted in circles with 4\arcsec\ radius.
The highest count number we found was 20\,counts, but the count statistics
suggests a Poissonian statistics with a peak at only 3.2 counts. With the
same search repeated for the chip in transformed coordinates we detected the
22\,counts at the edge of the ring system and only one further denser region
with 20\,counts at x$=12.5$\arcsec\ and y$=-70$\arcsec. After this exercise
we regard the photon excess at the edge of the ring system as significant, but
we have no suggestion for a production mechanism. It is not clear why this
excess is seen on only one side of the ring system. We checked the moons, but
none was near that particular position at any time during the
observation\footnote{http://ringmaster.arc.nasa.gov}. Also, a background
source is improbable, because we did not find the source on the
non-transformed detector coordinate system. The "light curve" of these
photons shows no significant anomaly, such that no instrumental effect can be
held responsible.\\

\begin{figure}[!ht]
 \resizebox{\hsize}{!}{\includegraphics{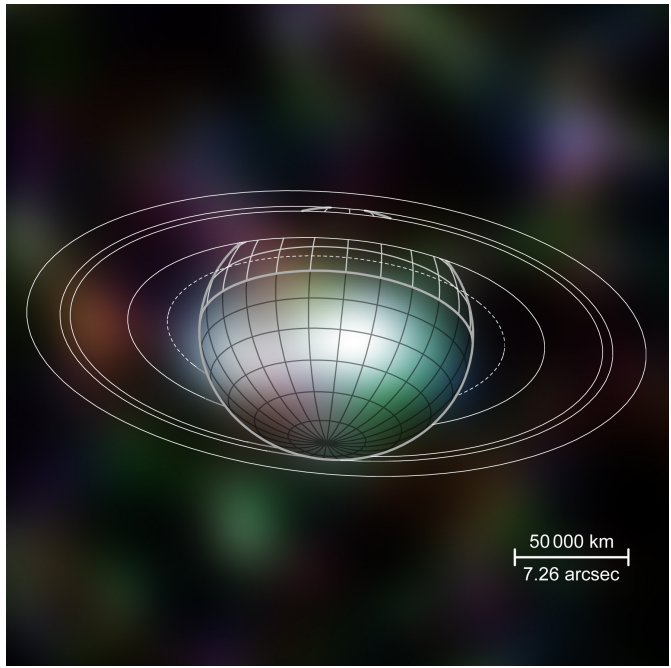}}
\caption{\label{vbild}X-ray colour LRGB image of Saturn in the energy ranges
0.4--0.6\,keV (red), 0.6--0.8\,keV (green), 0.8--1.0\,keV (blue), smoothed
using a Gaussian with a FWHM of 5\,arcsec. A drawing of Saturn
at the time of the observation, obtained from
http://ringmaster.arc.nasa.gov/tools/viewer2\_satc.html
was overlaid for clarity.}
\end{figure}

\subsection{Lightcurve}
\label{lightcurve}

\begin{figure}[!ht]
  \resizebox{\hsize}{!}{\includegraphics{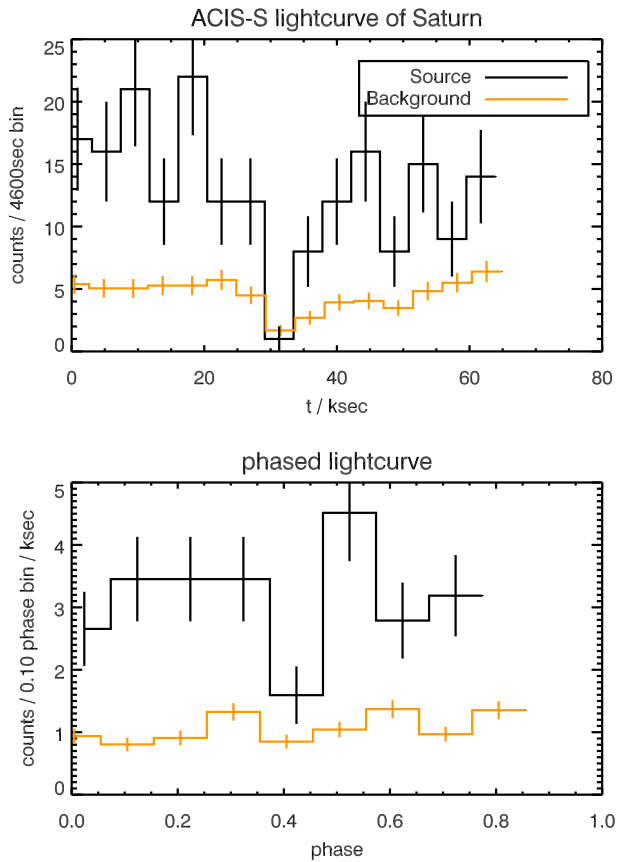}}
\caption{\label{lc}Light curve with time bins 4600\,sec (upper panel) and phased
lightcurve obtained from Saturn's rotation period 37.67\,ksec (=0.436\,days;
bottom panel) for source and background separately.}
\end{figure}

As described in Sect.~\ref{anal} we extracted the lightcurve and the spectrum
of the photons within 10\arcsec\ around the apparent position of Saturn.
We plot the lightcurve with a time binning of 4600\,sec in Fig.~\ref{lc}.
The dip at 35\,ksec is due to the gap between the two consecutive observations
with the Chandra spacecraft being repointed to follow Saturn's apparent
motion. For about 6\,ksec Saturn was actually outside the field of view.
The count statistics is obviously poor, however, there is a hint for two dips
possibly caused by rotational modulation. We therefore generated a phased
lightcurve using the known rotation period of Saturn (0.436\,days=37.67\,ksec)
shown in the bottom panel of Fig.~\ref{lc}; for Fig.~\ref{lc} a phase binsize
of 0.1 was used; note that because of the exposure history and the
rotation period of Saturn almost no coverage was obtained for the phase bin
between 0.8 and 1.0. In the phased up light curve there appears to be a
minimum between phases 0.35 -- 0.50. In order to assess the statistical
significance of this dip we generated random phased light curves with the same
number of photons as recorded from our Saturn observation, uniformly distributed
over the available phase space. We determined the number of photons recorded
in a phase interval $\Delta\phi = 0-0.15$ such that the actual number of
recorded photons is minimal. A comparison of those numerical experiments with
the numbers obtained for Saturn shows that there is a chance of about 15\% to
obtain a phase dip of the same strength as recorded for Saturn in a data set
with constant count rate. We therefore conclude that there is no evidence for
a rotationally modulated signal and data covering far more than two rotation
periods are required for statistically significant studies.

\subsection{Spectrum}

\begin{figure}[!ht]
  \resizebox{\hsize}{!}{\includegraphics{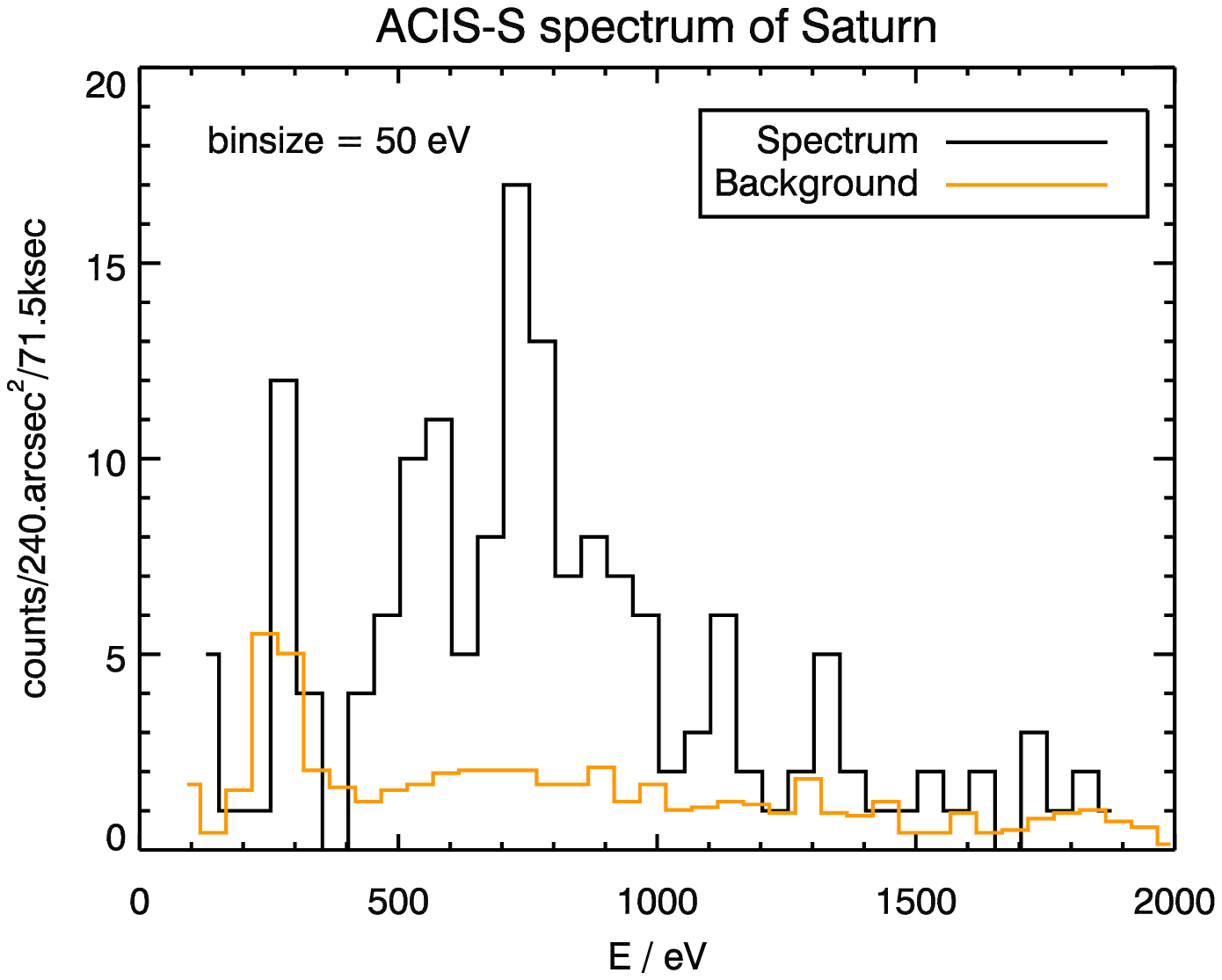}}
  \vspace{-.2cm}\resizebox{\hsize}{!}{\rotatebox{270}{\includegraphics{f6b}}}
\caption{\label{spec}Extracted spectra for the background and source+background
(top). Bottom: Best-fit obtained with XSPEC.
The model consists of a MEKAL (kT$=0.39\,\pm\,0.08$\,keV and solar abundances)
and a single line at 0.527\,keV, only instrumentally broadened.}
\end{figure}

The spectrum of the photons centered on Saturn's position is shown in
Fig.~\ref{spec}. The emission feature at $\sim 300$\,eV can be
identified both in the source and in the background.
The drop in counts just above 300\,eV corresponds to the carbon K-edge in
the optical blocking filter. The drop in counts at lower energies is also
due to absorption in this filter. Additional emission
features between 500\,eV and 800\,eV can only be identified for the source.
We tested a number of spectral models, and in the bottom panel of
Fig.~\ref{spec} we show the rebinned spectrum with our best-fit model, but we
point out that the number of photons is insufficient to arrive at firm
statements from the spectral modeling alone.\\

As a first approach we tested models assuming several single production
mechanisms (fluorescent scattering, solar wind charge exchange as observed in
comets, a powerlaw spectrum, thermal bremsstrahlung, black body, and a MEKAL
and Raymond-Smith spectrum, representing continuum plus line emission in
thermal equilibrium). We found no convincing proof for any of these models to
fully explain the X-ray emission from Saturn. Good fits were found with the
black body model (kT=0.18\,keV, $\chi^2_{\rm red}=0.7$ for 10 dof), but it
has no physical meaning. MEKAL and Raymond-Smith models yield only good fits
with adjusted elemental abundances. However, the spectral resolution does not
allow to claim abundance anomalies with high significance. If solar X-ray
emission was involved, we would expect to see the incident solar X-ray
spectrum with solar abundances.\\
In addition to single production mechanisms we considered a combination of
two mechanisms and obtained a good fit with a MEKAL model with solar abundances
combined with a single line from oxygen K-$\alpha$ ($\chi^2_{\rm red}=0.9$ for
9 dof). This fit is shown in Fig.~\ref{spec}. According to the best-fit we
find fluxes (0.1--2\,keV) of
$1.26\cdot10^{-15}\mbox{\,erg\,cm}^{-2}\mbox{\,s}^{-1}$ to be contained in the
fluorescence line and $5.5\cdot10^{-15}\mbox{\,erg\,cm}^{-2}\mbox{\,s}^{-1}$ in
the MEKAL model. The total flux in the energy interval 0.1--2\,keV derived
from the best fit is
$6.7\cdot10^{-15}\mbox{\,erg\,cm}^{-2}\mbox{\,s}^{-1}$, which is about 35\% of
the flux reported by \cite{ness00}.

\section{Discussion}
\label{disc}

With our 65.5\,ksec Chandra observation we clearly detect X-ray emission
originating from Saturn. The high spatial resolution allows to resolve
the spatial origin of the X-rays, the long exposure time covering two
rotational periods allows to find temporal anomalies, and the energy
resolution allows to extract a spectrum, but all analyses going
beyond the pure detection are difficult given the small number of photons.
Nevertheless we studied spatial, temporal, and spectral signatures to the
extent of qualitative statements, but quantitative analysis requires more
photons that can only be gathered with more observing time.\\

Our spectral modeling with XSPEC cannot give definitive answers for individual
production mechanisms. Additional hints come from the spatial distribution and
from comparison with other planets. We discuss potential production mechanisms
in the following.\\

\subsection{Single production scenarios}

{\bf Auroral X-ray emission} is expected to be concentrated towards the polar
regions as was found for Jupiter, while for Saturn we found no such
concentration in the southern polar region. However, this is required for
auroral emission, because Saturn's magnetic field is aligned with the rotation
axis. Also, UV auroral emission was found to be concentrated towards Saturn's
poles \citep{trau}. However, a
clear deficit of UV emission from the south pole is seen on recent HST
observations taken one month before our observations \citep{karosch}. The
model calculations by \cite{gilman86} are based on thick target bremsstrahlung
at high latitudes and their flux estimates are significantly below our measured
flux level. If the X-ray emission from the south pole reflects the
flux level predicted by \cite{gilman86}, we would not be able to isolate
the spectral signature from this emission in the spectrum. In our measurement
auroral emission is therefore not detected, but it cannot be excluded that
auroral X-ray emission originating only from the northern hemisphere was
occulted by the ring system.\\
Interestingly a {\bf Black Body} model with kT=0.18\,keV yields a good spectral
fit, but a physical meaning is difficult to find. The X-ray flux obtained with
this model is $f_X=4.4\cdot10^{-15}\mbox{\,erg\,cm}^{-2}\mbox{\,s}^{-1}$
in the energy interval 0.1--2\,keV.\\
{\bf Fluorescent Scattering} of solar X-rays in the upper atmosphere of Saturn,
the dominant process for the X-ray radiation observed from Venus and Mars,
would produce narrow emission lines from the most abundant elements in the
atmosphere. Elemental abundances are given by \citep{cameron82}: helium
(0.14\,H$_2$), oxygen ($1.4\cdot10^{-3}$\,H$_2$), carbon
($8\cdot10^{-4}$\,H$_2$), neon ($2\cdot10^{-4}$\,H$_2$), and nitrogen
($1.8\cdot10^{-4}$\,H$_2$). We found spectral models with up to four single
(narrow) emission lines discrepant from the spectrum. The oxygen fluorescence
flux measured for Mars was $5.4\cdot10^{-5}\mbox{\,ph\,cm}^{-2}\mbox{\,s}^{-1}$
\citep{mars}. With the assumption of equal physical conditions we can scale
this flux level to Saturn using the apparent diameter of Mars (20.3\arcsec)
and the distance (1.446\,AU). We thus expect fluorescent emission
from Saturn a factor 50 lower than for Mars, thus
$\sim 10^{-6}\mbox{\,ph\,cm}^{-2}\mbox{\,s}^{-1}$. When compared to the
Black-body photon flux ($8.7\cdot10^{-6}\mbox{\,ph\,cm}^{-2}\mbox{\,s}^{-1}$)
this implies that
only about 10\% of the total flux can be caused by fluorescent scattering.\\
{\bf Power Law and Thermal Bremsstrahlung} can be excluded from the spectral
modeling and the expected spectrum from {\bf Solar Wind charge exchange}
is too soft to explain the measured spectrum.\\
{\bf Backscattering of solar X-rays}, as suggestive from the spatial
distribution, would result in an X-ray spectrum resembling the incident solar
spectrum. Raymond-Smith and MEKAL models yield good fits, but only with
adjusted abundances. When assuming the X-ray luminosity of the Sun to be
$L_X=2\cdot10^{27}$\,erg/s (which is a high figure), an albedo for Saturn
can be estimated from the measured X-ray flux from Saturn and the distance
between the Sun and Saturn (9\,AU; Table~\ref{tab1}). With the flux obtained
from the blackbody model,
$f_X=4.4\cdot10^{-15}\mbox{\,erg\,cm}^{-2}\mbox{\,s}^{-1}$, we find an albedo of
$>5.7\cdot10^{-4}$, which appears to be quite high. \cite{schmitt91} measured
X-ray emission from the Moon ($L_X=7.3\cdot10^{11}\mbox{\,erg}\mbox{\,s}^{-1}$),
and they reported clear evidence for scattering of solar X-rays. The albedo of
the moon is then calculated to $10^{-5}$. Although elastic backscattering of
solar X-rays is suggestive from the spatial distribution, it is not consistent
with the solar spectrum, and the albedo would have to be significantly higher
than for the moon.\\

\subsection{Combination of scatter processes}

Since no definitive results were obtained from considering isolated processes,
we modeled the spectra of two possible scenarios operating in conjunction with
each other. We modeled the spectrum resulting from backscattering (resulting
in a MEKAL spectrum with solar abundance) and from fluorescence of the most
abundant element besides H and He: oxygen (narrow emission line at 527\,eV).
The combined spectrum yields a good fit with a temperature of
kT$=0.39\,\pm\,0.08$\,keV in the MEKAL model (Fig.~\ref{spec}, bottom). The
formal photon flux in the fluorescence line is only 10\% of the total flux
($1.9\cdot10^{-6}\mbox{\,ph\,cm}^{-2}\mbox{\,s}^{-1}$). This is consistent with
the expected fluorescent flux level scaled from Mars, but the remaining 90\% of
the emission is yet to be explained.\\
From the combined spectral model we calculate an X-ray luminosity of
$8.7\cdot10^{14}\mbox{\,erg}\mbox{\,s}^{-1}$. This can be compared to Jupiter's
equatorial emission reported to be $3.6\cdot10^{15}\mbox{\,erg}\mbox{\,s}^{-1}$
\citep{waite96}, a factor 4.1 higher than Saturn's X-ray luminosity. When
scaling for different distance and diameters (Jupiter's distance 5.4\,AU and
diameter 143000\,km and Saturn's diameter 120500\,km), we expect a factor 3.9
higher luminosity for Jupiter's equatorial regions, which is consistent with
our measured model flux. The similar luminosity levels suggest similar
production mechanisms for Jupiter's non-auroral emission and Saturn's total
emission. While \cite{waite97} explain the equatorial emission by heavy ion
precipitation, \cite{maur00} modeled two alternative mechanisms for low-altitude
X-rays and found solar photon scattering (90\% elastic scattering and 10\%
fluorescent scattering) consistent with ROSAT measurements. However, their
models predict non-auroral luminosities ($3\cdot10^7$\,W) about a factor 10
below the power output derived from the observations
\citep[$3.6\cdot10^8$\,W;][]{waite96}. This is in line with our considerations
about the required albedo for scatter processes. It would be an interesting
finding if the albedo of the gas giants Jupiter and Saturn were so much higher
than the albedo for the moon.

\section{Summary and Conclusions}
\label{concl}

 X-ray emission from solar system objects has revealed a large variety of
production mechanisms ranging from reflection of solar X-rays (Moon), solar
wind charge exchange in comets, fluorescent scattering of solar X-rays in
Venus and Mars to magnetically induced auroral emission in Earth and Jupiter.
Since Saturn is a gas planet with a magnetic field and auroral UV emission
\citep[e.g.,][]{ball,trau,bhar}, a certain level of X-ray emission has
always been expected. We carried out a Chandra observation of Saturn with the
intention to unambiguously detect X-ray emission from Saturn. A concentration
towards the poles would have been easier to detect, but with our observation
settings we were able to establish an unambiguous detection of X-ray emission
from Saturn, although no strong spatial concentration of X-ray photons was
found. A smoothed image of Saturn with a drawing of Saturn at the time of the
observation is shown in Fig.~\ref{vbild}. The production mechanism(s) for the
detected X-ray
emission cannot clearly be allocated from the available data. Possibly a
combination of several processes must be considered, but due to the limited
spectral signature we focused only on some possible scenarios as isolated cases.
We essentially found no convincing case, but the combination of two scatter
mechanisms, namely, elastic backscatter and fluorescent scattering of solar
X-rays, yield good spectral fits, are consistent with the detected spatial
distribution, but require an albedo a factor 50 higher than for the moon.\\

 Clearly, the X-ray production mechanisms for Saturn are different from those
for Jupiter. Not only the X-ray flux is significantly lower than
Jupiter's flux, but also the spatial distribution appears to be different. A
concentration towards the poles as encountered for Jupiter suggests magnetic
fields to play an important role, which is the case in most intrinsic X-ray
production mechanisms as, e.g., the solar corona. From two ROSAT HRI
observations of Jupiter in 1992 and 1994 more auroral X-ray emission from the
northern hemisphere than from the south pole was found \citep{waite96}. This
asymmetry is also seen in UV emission \citep{liven91}. A correlation of auroral
activity with infrared emission from H3$^+$ was reported by \cite{jupinfra}
for Jupiter and by \cite{satinfra} for Saturn with variable infrared fluxes
from both poles. From our X-ray observation we have no evidence at all for
auroral emission from Saturn, however, the northern polar region was
occulted by the ring system. If the north-south asymmetry phenomenon found for
Jupiter applies to Saturn as well, more emission with a soft
signature is expected to be measured when the north pole is not eclipsed. Such
a view was observed by ROSAT, but the detection was so marginal that additional
observations with Chandra are necessary to make a better case.\\

 We found our emission level consistent with equatorial emission from Jupiter
reported by \cite{waite96}, and since our X-ray emission is concentrated
towards the central part of the apparent disk the same production mechanisms
are anticipated. \cite{bhar02} point out that elastic scattering ($\sim 90$\%)
in conjunction with other processes can easily account for Jupiter's non-auroral
emission and should not be underestimated. Our spectral models are consistent
with this scenario and the spatial distribution is suggestive of scatter
processes, but a high albedo is required for scatter scenarios both for
Jupiter's equatorial X-ray emission and for the detected X-ray emission from
Saturn. We suspect more processes to operate in conjunction with scatter
processes producing the high level of X-ray emission at low latitudes.
Additional auroral emission at high northern latitudes cannot be ruled out from
our observation, but could only be discovered when the northern hemisphere is
not eclipsed.


\begin{acknowledgements}
J.-U.N. acknowledges support from DLR under 50OR0105.

\end{acknowledgements}

\bibliographystyle{aa}
\bibliography{jn,sat,jhmm}

\end{document}